# Three octave spanning supercontinuum by red-shifted dispersive wave in photonic crystal fibers

**Mohit Sharma\*and S. Konar**

*Department of Physics, Birla Institute of Technology, Mesra*
*\*Corresponding author: mohitsharmac@gmail.com, mohit@bitmesra.ac.in*

**Abstract:-**This paper presents a three layer index guided lead silicate (SF57) photonic crystal fiber which simultaneously promises to yield large effective optical nonlinear coefficient and low anomalous dispersion that makes it suitable for supercontinuum generation. At an operating wavelength 1550 $nm$, the typical optimized value of anomalous dispersion and effective nonlinear coefficient turns out to be $\sim 4\ ps/km/nm$ and $\sim 1078\ W^{-1}km^{-1}$, respectively. Through numerical simulation it is realized that the designed fiber promises to exhibit three octave spanning supercontinuum from 900 to 7200 nm by using 50 fs 'sech' optical pulses of 5 kW peak power. Due to the cross-phase modulation and four-wave mixing processes, a long range of red-shifted dispersive wave generated, which assist to achieve such large broadening. In addition, we have investigated the compatibility of supercontinuum generation with input pulse peak power increment and briefly discussed the impact of nonlinear processes on supercontinuum generation.

## 1. Introduction

Since the discovery in 1996 by Knight et al. [1], photonic crystal fibers have remain a very actively researched topic due to wide applications of these fibers in optical communications, signal processing and fiber optic devices [1-10]. In a photonic crystal fiber (PCF), tiny air holes are arranged in a suitable configuration across the cross section of the fiber which run along the length of the fiber. In a PCF optical properties can be tailored to achieve desired value by creating an appropriate air hole pattern and changing hole diameter, hole pitch and number of air hole rings. By appropriately changing these parameters, it is possible to achieve single modeness over large wavelength range, uniform dispersion [2-4], high as well as low nonlinearity [3-6], large birefringence [6-10] etc.

In a PCF large nonlinearity can be realized by making a very tight confinement of the optical mode in the core region, which could be achieved either by decreasing the core area with some modification in the cladding design or increasing the refractive index contrast between core and cladding by introducing soft glass materials with large nonlinear coefficient. Due to large nonlinear coefficient of non-silica materials, several authors have investigated PCFs which are made of these materials. For example, PCFs made of fluoride glass, bismuth oxide, lead silicate, tellurite and chalcogenide etc., exhibit large effective nonlinearity which is suitable for devices that rely on optical nonlinearity. Particular interest has been shown to lead silicate glass with nonlinear coefficient which is 20 times larger than that of silica glass and has good thermal and crystallization stability [11, 12]. Lead silicate glass fibers exhibit attenuation ~1-3 dB/m at the operating wavelength 1.55 $\mu m$, whereas chalco-sulfide glass has fiber attenuation 3-4 dB/m at 1.55 $\mu m$ [12] and it has higher thermal expansion coefficient than silica, which may allow more flexibility in the fabrication process. A brief comparison of thermal, crystalline and other properties of soft glasses have been discussed by Feng et al. [12]. PCFs with large effective optical nonlinearity require very low threshold power to generate broadband supercontinuum, which make them attractive for application in supercontinuum (SC) generation.

SC generation in PCFs has been widely studied by several researchers [6-46] during last fourteen years. SC generation is a complex process of spectral broadening of nanosecond [13, 14], picosecond [15, 16] or femtosecond [17, 6] optical pulses, which undergo through a number of nonlinear optical interactions like self-phase modulation (SPM) [18], modulation instability (MI) [16, 19], stimulated Raman scattering (SRS) [18], four-wave mixing (FWM) [13, 15, 18, 19], self-steepening, soliton fission [19] etc., in an optical nonlinear medium such as a PCF. These nonlinear processes are governed by pulse duration, pumping wavelength and input peak power, whereas the group velocity dispersion (GVD) and its higher order terms at the pumping wavelength play a vital role in determining the quality of the continuum and its shape. To achieve broadband supercontinuum, the high power input pulse is generally launched near the zero dispersion point in a highly nonlinear fiber. PCFs with large effective nonlinearity require very low threshold power to generate the wideband SC, on the other hand low and uniform dispersion enables four-wave matching leading to wideband flat spectra. Thus, large effective nonlinearity and low uniform dispersion profile is the key to broadband flat SC spectra. When an optical beam propagates faster than the phase velocity of light in a dispersive medium then some distorted wave generated called dispersive wave (DW). These DW could be either blue-shifted [6] if third order dispersion (TOD) is positive or red-shifted if the TOD is negative [22]. The blue-shifted DWs holds much attention of the scientific communities. The ample number authors have generated a blue-shifted DWs like, A range of DW was observed from 580 to 630 nm by employing 300-mW ytterbium laser [23], from 300 to 800 nm by using an erbium laser with 7% energy transfer efficiency [24], nearby 430 nm with 800-nm pumping with 15% of the total energy [25], nearby 640 nm with a fifth order soliton[26] and from 1626 to 1685 nm with the average pump power changing from 250 to 440 mW at the pump wavelength of 1760 nm [27]. The red-shifted DW depended mainly on ultra-short pulses with a verylow anomalous dispersion at the pumping wavelength. This provides the multiple solitons generation due to which the XPM and FWM takes place and generate the red shifted DW. In anteceding decade, several studies has been performed in order to achieve red-shifted DW. The red-shifted DW was observed approximate 1350 nm in multimode fiber with pumping wavelength 1064 [28], around 2100 nm with pumping wavelength 1500 nm [29] and from 1240 nm to 1400 nm with containing 76.7 % of the total energy at the pumping wavelength 1064 nm [30]. Whereas, in the continuing work, we have reported approximately 6300 nm DW wave from 900 to 7200 nm with a pumping wavelength 1550 nm.

Due to large effective nonlinearity, PCFs made of SF57 glass have remain very attractive nonlinear medium for SC generation. Extensive investigations have been carried out towards evolving designs of SF57 glass PCFs to achieve large effective nonlinearity and small dispersion so as to get flat broadband supercontinuum in these fibers. Along this direction, several authors have studied effective nonlinear co-efficient of SF57 glass PCFs. For example, effective nonlinear coefficient of 112, 500 and 1860 $W^{-1}km^{-1}$ at 1.55 $\mu m$ operating wavelength were respectively reported by Xing-Ping et al. [7], Tiwari et al. [8] and Leong et al. [10]. Several efforts have been made to achieve wideband SC spectra in SF57 glass PCFs [7-10]. For example, Xing-Ping et al. [7]

generated SC spectra from 1300 nm to 1900 nm, Leong et al. [10] were able to generate SC spectra spanning more than 1000 nm by using 300 fs pulses in a 6.8 cm long fiber. Miret et al. [9] have reported 3 dB flat SC spectra spanning over ~1500 nm by using femtosecond pulse in a 15 cm PCF which exhibits normal dispersion. Buczynski et al. [11] have also reported about 1500 nm wide SC spectra in lead-bismuth-galate glass PCF by pumping femtosecond pulses at operating wavelength 1540 nm.Tiwari et al. [8] generated SC spectra from 1000 nm to 3200 nm by using 50 $fs$ pump pulse of peak power 2 kW in a 15 cm long PCF. Along this direction, we focus our attention to achieve broadband SC generation at low input power. The objective of the present study is to two-fold. First, to design a SF57 glass PCF which exhibits large nonlinearity and low dispersion and then we use this fiber to achieve broadband SC spectra employing pump pulses with low peak power.The proposed fiber can be fabricated without much difficulties since several authors have demonstrated the fabrication of PCFs with soft glass. For example, Leong et al. [10] have fabricated a soft glass PCF with core diameter 0.95 $\mu m$, Xing-Ping et al. [7] have studied SF57 PCF with small air hole of diameter 0.5 $\mu m$ which was fabricated by Institute of Photonics and Advance Sensing at University of Adelaide, Australia.Buczynski et al. [11] have also demonstrated the fabrication of lead silicate PCF with core diameter 3.36 $\mu m$ by using extrusion technique [47].

The broadband SC sources spanning from 900 nm to 7200 nm are extremely useful and have chemical, biological, medical, and astronomical applications. They can be used for studying the composition of extrasolar planets and detecting the volatile compounds of human exhale air for diagnosis [48-51].

The organization of the paper is as follows: In section 2, we have highlighted the necessary computational procedure and essential fiber parameters. In section 3, we have described the PCFs and important results have been discussed in this section. In section 4, we have described a theoretical model of SC generation and discussed the results. A brief conclusion has been added in section 5.

## 2. Theoretical Model

Several numerical modeling techniques have been employed to characterize PCFs, including multipole method [31], localized-function method [32], plane-wave expansion method [33], finite element method [34, 35], approximate effective index model [36], finite difference frequency domain (FDFD) method [37] and finite difference time domain (FDTD) method [38, 39]. Gallagher [40] have presented the comparison study of numerical modeling in detail. In this paper, we use FDD technique to analyze the optical properties of photonic crystal fiber.

Fiber dispersion is one of the most important parameter relevant to supercontinuum generation. The total dispersion $D_C(\lambda)$ of the fiber is the sum of waveguide dispersion and material dispersion i.e., $D_C(\lambda) = D_M(\lambda) + D_W(\lambda)$ and numerically expressed by $D_C(\lambda) = -\frac{\lambda}{c}\frac{d^2 n_{eff}}{d\lambda^2}$, where $\lambda$ is operating wavelength, $c$ is the speed of light, $n_{eff} = \frac{\lambda}{2\pi}\beta$ is effective refractive index of the fiber, $n$ is the refractive index of SF57 glass and $\beta$ is the propagation constant. The Sellmeier's equation $n^2(\lambda) = 1 + \sum_i \frac{B_i \lambda^2}{\lambda^2 - C_i}$ has been utilized to evaluate material dispersion [32], where $B_i$ and $C_i$ is the Sellmeier's coefficients of the material. For SF57 glass $B_1 = 1.8165127$, $B_2 = 0.42889364$, $B_3 = 1.07186278$, $\lambda_1 = 0.0143704198 \; \mu m$, $C_2 = 0.0592801172 \; \mu m$ and $C_3 = 121.419942 \; \mu m$. The effective nonlinear coefficient $(\gamma)$ of the fiber can be calculated using $\gamma = (2\pi \, n_2 / \lambda \, A_{eff}) \times 10^3 \; W^{-1} km^{-1}$, where $n_2 = 4.1 \times 10^{-19} \; m^2/W$ and $A_{eff}$ is an effective mode area of the fiber, which is defined [3-10] by $A_{eff} = \frac{(\iint_{-\infty}^{\infty} |E|^2 dx dy)^2}{\iint_{-\infty}^{\infty} |E|^4 dx dy}$. The single modeness of a PCF is characterized by the normalized V parameter whose effective value $V_{eff}$ can be expressed as $V_{eff} = 2\pi \frac{\Lambda}{\lambda} \sqrt{n_{core}^2 - n_{eff}^2}$, where $\Lambda$ is the hole pitch, $n_{core}$ is the refractive index of the core. The single mode cut-off for photonic crystal fiber is $V_{eff} \leq 4.1$ [3, 19]. The confinement loss is calculated by using finite element method (FEM) [34, 35], which is numerically represented as $CL = 8.686 \, k_0 Im(n_{eff})$, where $Im(n_{eff})$ is the imaginary part of the effective refractive index and $k_0 = 2\pi/\lambda$ is the wavenumber in free space [52].

## 3. Three Layer Fiber

At the outset, we plan to design a three layer SF57 glass PCF which will be employed to generate broadband SC spectra. Since the aim is to accomplish broadband SC generation using very short PCF at low input power, we evolve the design in such a way that the PCF exhibits large effective nonlinearity and simultaneously small anomalous dispersion. In addition, the zero dispersion point of the fiber should be very close to the pump wavelength $(\lambda = 1.55 \; \mu m)$.

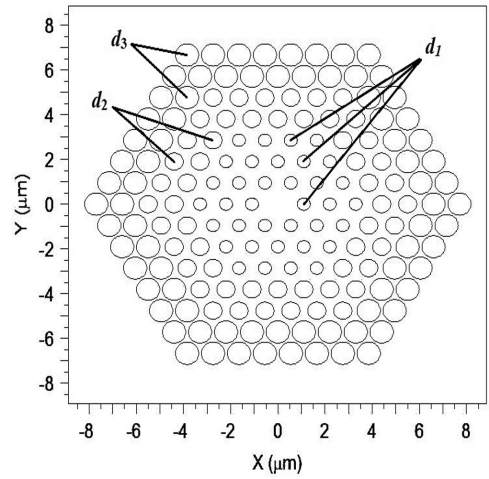

Fig.1 Transverse mode profile of proposed Fiber, where $d_1 = 0.5 \times \Lambda \; \mu m$, $d_2 = 0.7 \times \Lambda \; \mu m$, $d_3 = 0.9 \times \Lambda \; \mu m$.

The cross-section of the typical three layer unique PCF has been displayed in Fig. 1. Seven rings of air holes are arranged in the cross section of the fiber which are running along its length. The key to achieve optimized dispersion and nonlinearity lies in gradual increase in air hole diameter. Therefore, we have introduced air holes of three different size i.e., a three layer air hole structure. The diameter $(d_1)$ of air holes in three inner rings is $d_1 = 0.5 \times \Lambda$, the diameter $(d_2)$ of air holes in the fourth and fifth rings is $d_2 = 0.7 \times \Lambda$, while the diameter $(d_3)$ of air holes in the last two rings is $d_3 = 0.9 \times \Lambda$, where $\Lambda$ is the air hole pitch. The last two air hole rings of the fiber help to confine the optical mode tightly in the core, while rings of smaller air holes ensure low and uniform dispersion. In our subsequent discussion it will be evident that this differential increment of air hole diameter, as we move outward, ensure optimized value of large effective nonlinearity and low dispersion.

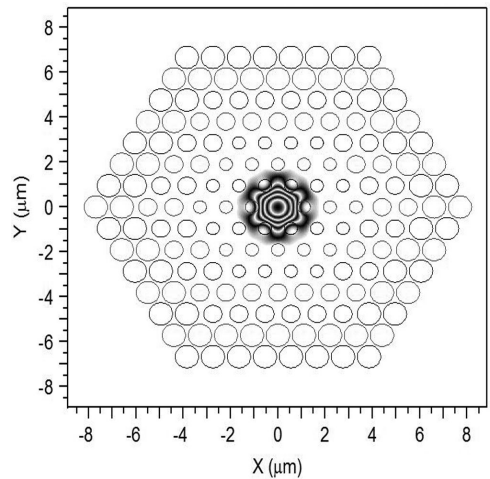

Fig.2 Contour map of transverse index profile at the operating wavelength 1.55 μm; $d_1 = 0.5 \times \Lambda \; \mu m$, $d_2 = 0.7 \times \Lambda \; \mu m$, $d_3 = 0.9 \times \Lambda \; \mu m$.

We now proceed to study the optical properties of the designed fiber. Typical fundamental mode field of the fiber at a wavelength $\lambda = 1.55 \; \mu m$ has been depicted in Fig. 2. The optical mode of the fiber is tightly confined in the core region. In order to study dispersion, we first calculate the effective refractive index of the fundamental mode of the fiber whose variation with operating wavelength $\lambda$ has been displayed in Fig.3. The effective refractive index decreases with the decrease in the value of hole pitch$\Lambda$. This is quite obvious since closer air hole spacing will reduce index contrast between core and cladding leading to reduction in effective refractive index. In order to have an idea about what happens when all the air holes are of same size, we have examined the effective refractive index of such fibers. In Fig. 3, the solid line represents the refractive index variation in a fiber whose air holes are of equal size, and this fiber has been designated as simple fiber. Note that effective refractive index for this particular case is much lower in comparison to the cases where air holes are of different size. In addition, the variation of refractive index in the present case is much larger. This larger variation of effective refractive index yields large dispersion, whereas slower variation of effective refractive index for unequal hole diameter leads to a smaller dispersion.

The variation of dispersion with wavelength of the designed fiber has been demonstrated in Fig. 4 for three different values of hole pitch, namely $\Lambda = 1.1, 1.2$ and $1.3\ \mu m$. Size of three different types of air holes are $d_1 = 0.5 \times \Lambda$, $d_2 = 0.7 \times \Lambda$and $d_3 = 0.9 \times \Lambda$. In the same figure, using a solid line we have also displayed the variation of dispersion exhibited by a fiber which possess same number of air holes of equal size and albeit large.Obviously, dispersion exhibited by this fiber is quite large in comparison to the fiber with unequal hole diameter. From figure, it is evident that the dispersion of the fiber reduces significantly when size of the hole diameter in different rings in not equal. Particularly, significant decrease in the dispersion is achievable for three layer arrangement in which three innermost rings possess air holes of smallest diameter, two intermediate rings possess air holes of slightly larger diameter and the two outermost rings possess air holes of largest diameter. This three layer arrangement of air holes ensure small dispersion, and subsequently we will demonstrate that this will also ensure large nonlinearity. The zero dispersion wavelength of this fiber can be shifted by increasing the hole pitch $\Lambda$. For a hole pitch $\Lambda = 1.1\ \mu m$, the fiber possess two zero dispersion point, one at $\sim 0.35\ \mu m$ while the other at $1.57\ \mu m$. These parameter could be useful for generation of supercontinuum with a pump at $1.55\ \mu m$.

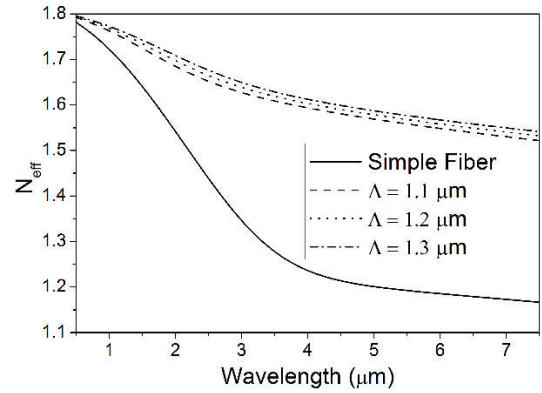

Fig. 3 Effective refractive index of the fiber with $d_1 = 0.5 \times \Lambda\ \mu m$, $d_2 = 0.7 \times \Lambda\ \mu m$, $d_3 = 0.9 \times \Lambda\ \mu m$; Solid line represents a simple fiber in which all air holes are of equal diameter $d = 0.9 \times \Lambda\ \mu m$with air hole pitch $\Lambda = 1.1\ \mu m$.

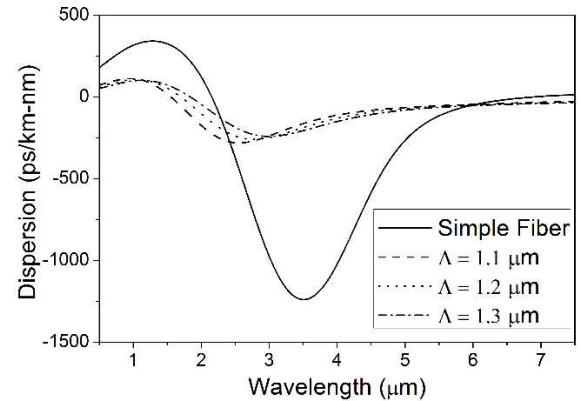

Fig. 4 Dispersion profile of the fiber. $d_1 = 0.5 \times \Lambda\ \mu m$, $d_2 = 0.7 \times \Lambda\ \mu m$, $d_3 = 0.9 \times \Lambda\ \mu m$. The solid line demonstrates dispersion of a typical PCF whose air holes are of same size ($d = 0.9 \times \Lambda\ \mu m$) with air hole pitch $\Lambda = 1.1\ \mu m$.

We now proceed to investigate effective nonlinear coefficient of the fiber. To do this, we first examine the fundamental mode of the fiber. The variation of effective area of the fundamental mode with wavelength has been depicted in Fig. 5(a) for three different values of hole pitch $\Lambda$, especially $\Lambda = 1.1, 1.2$ and $1.3\ \mu m$. To make the present fiber design more convincing, like the previous figure, we have displayed in the same figure, with solid line, variation of effective mode area of the fiber which possess same number of air holes of equal diameter $d = 0.9 \times \Lambda$.When all air holes are of equal size, the core cladding index contrast is large which leads to tight confinement of the optical mode in the core and hence effective mode area of this fiber is very small. Consequently, a fiber with equal air hole diameter will exhibit large nonlinearity. Inspite of this, we ignore this particular design with identical air holes since it also exhibits large dispersion which is detrimental to SC generation. Therefore, we turn our attention to the fiber which possess air holes of different diameter. From Fig. 5(a) it is evident that effective area of the fundamental mode gradually increases with the increase in the wavelength. It also increases with the increase in$\Lambda$. The variation of effective nonlinearity with the wavelength has been demonstrated in Fig. 5(b). As anticipated, the designed fiber exhibits large effective nonlinearity which decreases with the increase in wavelength. Note that the solid line in the Fig. 5(b) depicts the effective nonlinearity of the fiber whose air holes are of the equal diameter. As pointed out earlier, though this particular fiber exhibits large nonlinearity it is not suitable for SC generation since it also exhibits large dispersion.

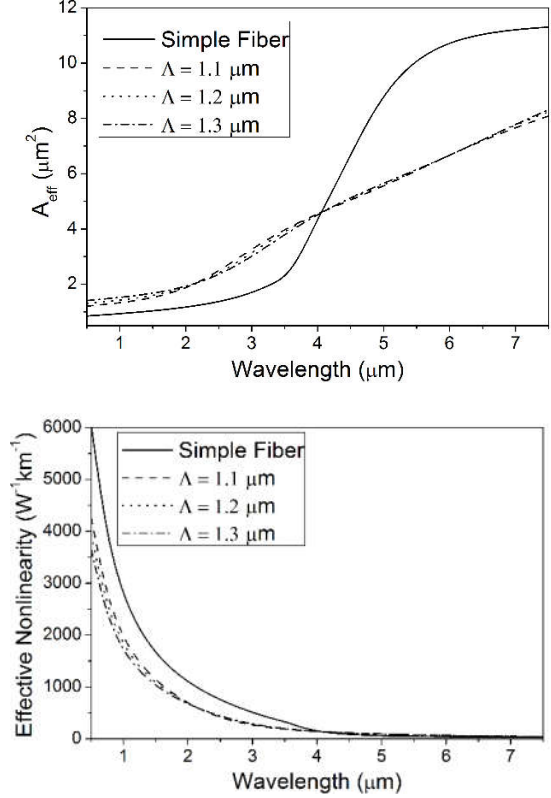

Fig. 5 (a) Effective optical mode area, (b) effective nonlinear coefficient at different hole pitch with $d_1 = 0.5 \times \Lambda$ μm , $d_2 = 0.7 \times \Lambda$ μm , $d_3 = 0.9 \times \Lambda$ μm. Solid line represents simple fiber in which all air holes are of equal diameter d = $0.9 \times \Lambda$ μm with air hole pitch $\Lambda = 1.1$ μm.

We now proceed to find out a suitable value of hole pitch which will lead to a fiber that exhibits sufficiently large nonlinearity as well as low dispersion. To find out the optimized value of nonlinearity and dispersion, we have displayed in Fig.6 the interdependence of dispersion and effective nonlinearity of the fiber for different values of Λ. Point A in the graph represents the fiber whose air holes are of same size. The fiber possesses large nonlinearity as well as large dispersion. From the figure it is amply clear that, a fiber with $\Lambda = 1.1$ μm yields sufficiently large nonlinearity and simultaneously exhibits low dispersion (point C in the figure). At the wavelength 1.55 μm, typical values of effective nonlinearity and dispersion are $\sim 1078 \ W^{-1}km^{-1}$ and $\sim 4 \ ps/km/nm$ respectively. Therefore, this fiber could be useful for supercontinuum generation with a pump at 1.55 μm.

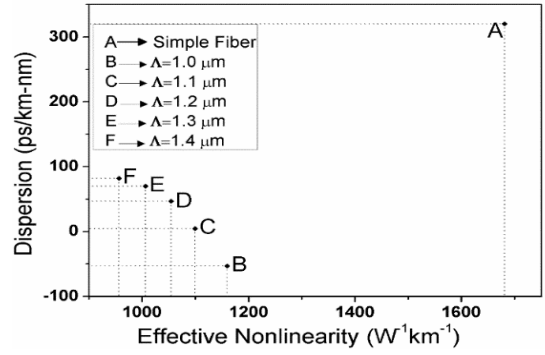

Fig. 6 Dispersion and effective nonlinear coefficient relation at the operating wavelength 1.55 nm with fiber parameter $d_1 = 0.5 \times \Lambda$ μm, $d_2 = 0.7 \times \Lambda$ μm , $d_3 = 0.9 \times \Lambda$ μm. Simple fiber in which all air holes are of equal diameter d = $0.9 \times \Lambda$ μm with air hole pitch $\Lambda = 1.1$ μm.

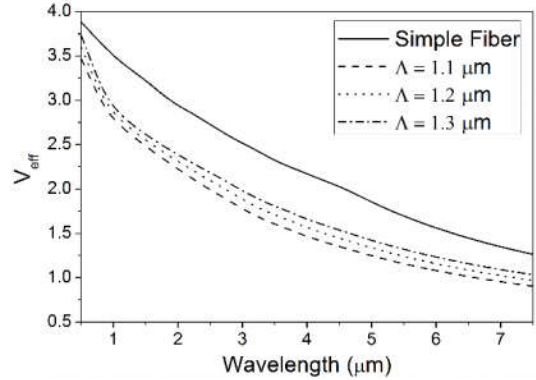

Fig. 7 Variation of effective V-parameter with wavelength. $d_1 = 0.5 \times \Lambda$ μm , $d_2 = 0.7 \times \Lambda$ μm , $d_3 = 0.9 \times \Lambda$ μm. Solid line represents a simple fiber in which all air holes are of equal diameter d = $0.9 \times \Lambda$ μm with air hole pitch $\Lambda = 1.1$ μm.

In order to examine the single modness of the fiber, we have demonstrated the variation of effective V-parameter with wavelength in Fig. 7. The value of $V_{eff}$ gradually decreases with the increase in the value of wavelength. From the figure it is evident that $V_{eff} \leq 4.1$, [3, 19] hence, the fiber is endlessly single mode over a wide range of wavelength.

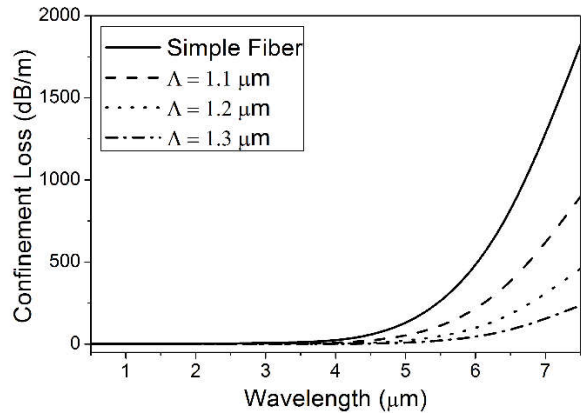

Fig. 8 Variation of confinement loss with wavelength. $d_1 = 0.5 \times \Lambda$ μm, $d_2 = 0.7 \times \Lambda$ μm , $d_3 = 0.9 \times \Lambda$ μm. Solid line represents the confinement loss of a simple fiber in which all air holes are of equal diameter d = $0.9 \times \Lambda$ μm with air hole pitch $\Lambda = 1.1$ μm.

To this end, it is worth investigating the confinement loss of the fiber. It would be also appropriate to compare the confinement loss of the designed fiber with that of the simple fiber. In Fig.8, we have demonstrated the variation of confinement loss of the fiber with wavelength. It is amply clear that initially our designed fiber as well as the simple fiber show very low confinement loss upto 5 $\mu m$. Above 5 $\mu m$ confinement loss increases with the increase in wavelength. Notice that, the increase in confinement loss of our designed fiber above 5 $\mu m$ is much less in comparison to that of the simple fiber. This unique behaviour also justifies the three layer structure evolved in this communication.

## 4. SC Generation in PCF

The pulse propagation in the PCF has been modeled [41, 42] employing generalized nonlinear Schrödinger equation:

$$\frac{\partial}{\partial z}A(z,T) = -\frac{\alpha}{2}A(z,T) + \sum_{n \geq 2}\beta_n\frac{i^{n+1}}{n!}\frac{\partial^n}{\partial T^n}A(z,T)$$
$$+ i\gamma\left(1 + \frac{i}{\omega_0}\frac{\partial}{\partial T}\right)A(z,T)$$
$$\int_{-\infty}^{\infty}R(T')\,|A(z,T-T')|^2 dT' \qquad (1)$$

where $A(z,t)$ is the envelope of the electric field of the input pulse, $\alpha$ is the fiber loss, $\beta_n = d^n\beta/d\omega^n$ is the nth order dispersion at the pumping wavelength $\lambda_P$. R is the nonlinear response function, which consists of an instantaneous electric response and a contribution from delayed Raman response. The Raman response function $R(T)$ may be defined as $R(T) = (1-f_r)\delta(T) + f_r h_r(T)$, where $f_r = 0.1$ for SF57 material and $h_r$ is calculated using $h_r(T) = \frac{\tau_1^2 + \tau_2^2}{\tau_1\tau_2^2}exp\left(\frac{-\tau}{\tau_2}\right)sin\left(\frac{\tau}{\tau_1}\right)$, with $\tau_1 = 5.5\ fs$ and $\tau_2 = 32\ fs$. The nonlinear Schrödinger equation has been solved by using the *split-step Fourier method* [32].We have taken 15 cm long fiber whose parameters are as follows: $\Lambda = 1.1\ \mu m$, $d_1 = 0.5 \times \Lambda$, $d_2 = 0.7 \times \Lambda$ and $d_2 = 0.9 \times \Lambda$. From the dispersion profile $D_c(\lambda)$ of this fiber that has been depicted in Fig.4, the value of total chromatic dispersion at a wavelength 1.55 $\mu m$ turns out to be $\sim 4\ ps/nm.km$. We have calculated values of $\beta_n$'s at 1.55 $\mu m$ wavelength which are $\beta_2 = -0.5703\ ps^2/km$, $\beta_3 = 4.897 \times 10^{-2}ps^3/km$, $\beta_4 = -1.898 \times 10^{-4}ps^4/km$, $\beta_5 = 1.9878 \times 10^{-7}ps^5/km$, $\beta_6 = 1.2423 \times 10^{-9}ps^6/km$, $\beta_7 = -6.4076 \times 10^{-12}ps^7/km$, $\beta_8 = -5.8448 \times 10^{-16}ps^8/km$, $\beta_9 = 1.5628 \times 10^{-16}ps^9/km$ and $\beta_{10} = 8.1982 \times 10^{-17}ps^{10}/km$. For numerical simulation, we have launched *sech* optical pulses at 1.55 μm wavelength in the 15 cm long PCF. The pulse duration and peak power of these pulses are 50 *fs* (FWHM) and 1 kW respectively. The Fiber loss may be neglected due to short length (15 *cm*) PCF and it is expected that the loss will not influence supercontinuum generation significantly.

The higher order solitons becomesgoverning phenomenon when ultra-short pulses pumped in the anomalous dispersion regime. The higher order solitons are determined by the soliton numbers [42] using $N^2 = L_d/L_{NL}$, where dispersion length $L_d = T_0^2/|\beta_2|$, nonlinear length $L_{NL} = 1/\gamma P_0$, N is the soliton number, $P_0$ is the peak power, $T_0$ is the temporal FWHM pulse duration and $\beta_2$ is the group velocity dispersion coefficient. The soliton numbers for the SCare calculated with respect to the variation of input pulse duration in Fig. 9, which reveals that decrement in the input pulse duration leads to the generation of large number of higher order solitons. The soliton numbers calculated as very large about $N = 87$ at the peak power 5 kW, as the second order dispersion coefficient reported very low value.This much of large solitons generation may leads to intra-soliton interaction, which is discussed in the next section. The calculated dispersion length $L_d = 141\ cm$of the designed fiber is much larger than the fiber length L = 15 cm. In that case, the dispersion effect become negligible in comparison to the nonlinear term and pulse evolution in the fiber is governed by the nonlinear effects that leads to spectral broadening of the pulse [42].As the soliton fission lengths are also important to the effect of pulse compression, where it attains maximum spectral bandwidth. The soliton fission lengths [42] are calculated by using $L_{fiss} = L_d/N$, which is turned out as $L_{fiss} = 1.6\ cm$ at the peak power 5 kW. These calculated values of soliton fission lengths are found to be agrees well with the simulation result of Fig. 11. The soliton fission lengths are depicted with the variation of input pulse duration in Fig 10. It reveals that the smallpulse duration helps in early pulse compression causes the small soliton fission length, resulted a wide spectrum are achieved at a very short length.

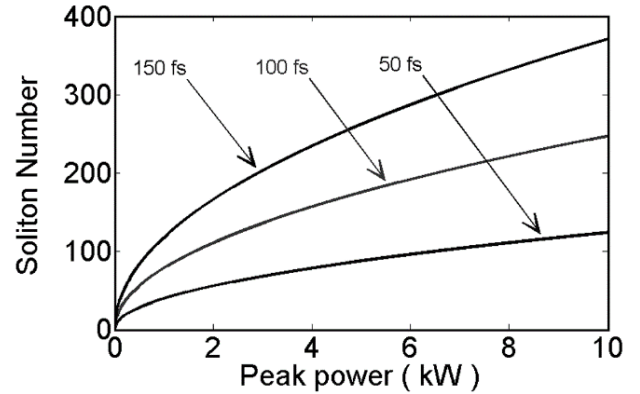

Fig. 9 Soliton numbers with the variation of input pulse duration.

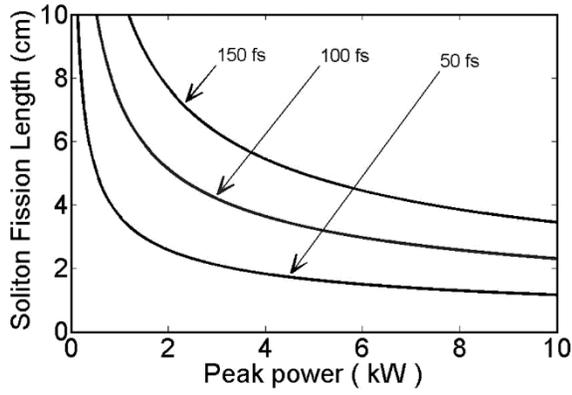

Fig. 10 Soliton fission lengths with the variation of input pulse duration.

The spectral and temporal evolution of the input pulse have been displayed for different length of the fiber in Fig. 11(a) and 11(b) respectively. It is evident from the figure that significant SC spectral broadening is achieved at the end of the fiber. Additional information of spectral and temporal broadening dynamics may be captured by using density plot. The density plot has been depicted in Fig. 12, in which both the spectral and temporal profile have been plotted using logarithmic density scale truncated at -40 dB relative to the maximum value. Such plot is very useful for study the low amplitude spectral and temporal component [2].

The initial broadening is led by SPM and maintained its influence till end of the fiber in blue regime. As, we have reported very low value of second order dispersion coefficient, the large number of higher order solitons generated at the fission length $L_{fiss} = 1.6\ cm$, Due to high pulse compression several new wavelength components in the trailing edge of the temporal profile (Fig. 11(b)), which confirms the existence of FWM and SRS effects. The weak SRS effect is dominated by the FWM process which generated a weak DW in blue regime and due to positive TOD, the DW are blue shifted with respect to the pumping wavelength [22]. At the fiber end, some peaks are observed in DW at the lower wavelength regime. According to Cristiani et al. [25], each DW peak present in a SC output corresponds to a higher order soliton excited by the input pulse. The spectrum in Fig. 11 (a) reveals that the DW peaks in the lower wavelength regime are the resembled of some higher order solitons. At the fiber end, these peaks are observed at the lower wavelength from 800 nm to 2000 nm in Fig. 11 (a).As the input pump energy $E_P = 2|\beta_2|/\gamma T_0$ is comparatively high for ultra-short pulse (for this study 50 fs), the XPM and FWM between the solitons and DW components [43-45] and the interactions between solitons themselves [46] influences the SC toward the longer wavelength. Initially, due to the interaction of multiple higher order solitons, the XPM induced DW offers the flat spectrum (at fiber length 6 cm in Fig. 11). Later on, DW generated by the influence of FWM in longer wavelength disturbed the SC flatness and start forming several peaks in the spectrum. And at the end of the fiber some distorted peaks observed, due to the combined effect of XPM and FWM processes. Aforementioned nonlinear processes offer the wide band SC generation from 800 nm to 7200 nm.

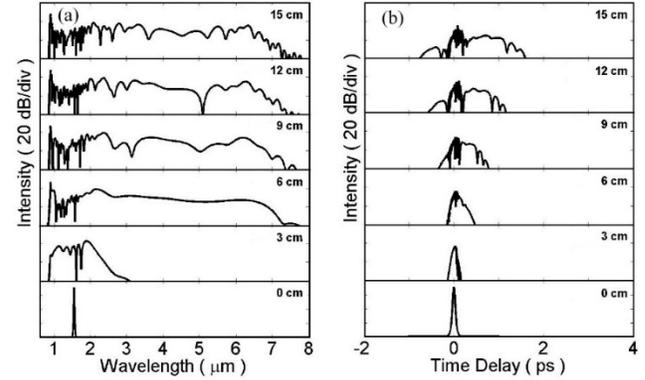

Fig.11 (a) Spectral (b) temporal profile of supercontinuum by pumping 50 fs pulse of 5 kW peak power in a 15 cm long PCF.

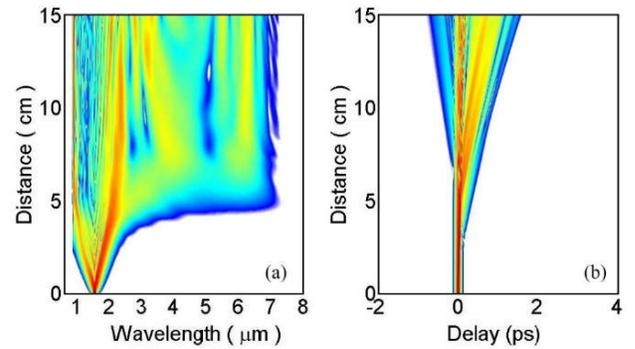

Fig. 12 Spectral and temporal evolution of supercontinuum generation by pumping 50 fs pulse of 5 kW peak power in a 15 cm long PCF.

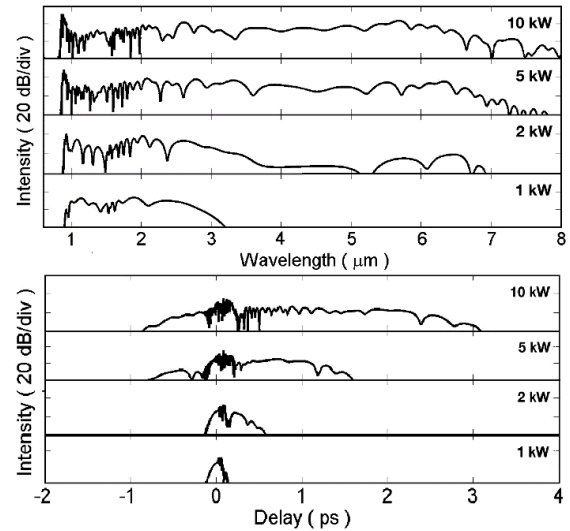

Fig. 13 Spectral and temporal evolution of supercontinuum generation by pumping 50 fs pulse in a 15 cm long PCF at different peak power.

In order to study the influence of input pulse peak power, we have plotted the spectral and temporal profile for different pump powers. It is evident from Fig. 13, the increase pump power offers XPM and FWM to generate the red-shifted DW. Moreover, the Fig. 14 reveals that the initial broadening is caused specially by SPM at the low pump power of 1kW. Later on, XPM generated red-shift of DW offers wide band spectral broadening. For the continuing pump power, the intra-solitons interaction introduces XPM effect which offers the smooth spectrum which turned out into distorted SC due to the FWM effect. At the conclusion of this section, the DW generation in blue regime are influenced solely by FWM while the DW generation in red regime are offered by the XPM due to intra-solitons interaction and if power increases the additional FWM generated the DW which disturbed the SC flatness.

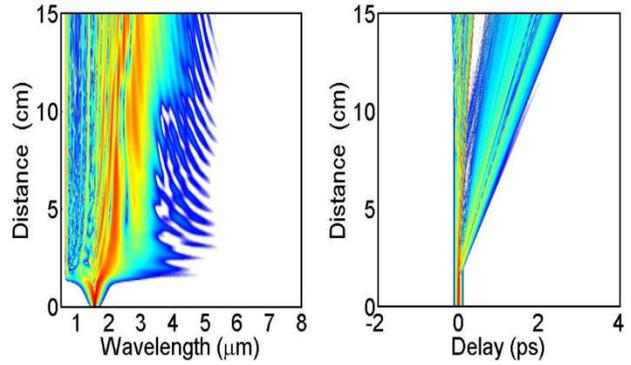

Fig. 15 Spectral and temporal evolution of supercontinuum generation by pumping 50 fs pulse of 5 kW peak power in a 15 cm long PCF.

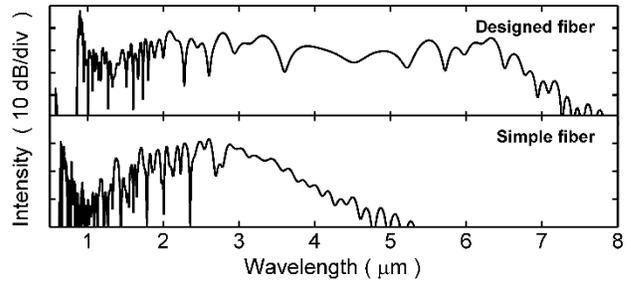

Fig. 16 Supercontinuum spectra generated at the end of designed and simple fibers using 50 fs pulses of 5 kW peak power. Length of both fibers 15 cm.

At this stage, it would be worth comparing the performance of SC generation by the designed fiber with that of the simple fiber. For comparison we have taken equal fiber length of 15 cm. In both fibers we have injected optical pulses of 50 fs duration and 5 kW peak power. Fig.15 demonstrates the SC generation in the simple fiber. In order to compare the quality and broadness of the SC spectra generated by these two fibers, we have depicted these spectra inFig.16. From Fig.16 it is evident that designed fiber is capable of producing a much broader spectra. For example, while in the simple fiber SC extends upto 4700 nm, it extends upto 7200 nm in the designed fiber. Though optical nonlinearity exhibited by the simple fiber is large, yet the SC spectra generated by it is much smaller in comparison to that of designed fiber. Note that at the pump wavelength simple fiber exhibits larger optical nonlinearity ($1750\ W^{-1}km^{-1}$)in comparison to that of designed fiber ($1078\ W^{-1}km^{-1}$). Therefore, the wider SC spectra generated in the designed fiber is attributed to lower value of dispersion in the designed fiber whose value is4 $ps/km/nm$ at the pump wavelength while it is 341 $ps/km/nm$ in the simple fiber at same wavelength. Moreover, the designed fiber exhibits flat SC over a wide range of wavelength in comparison to the simple fiber. This also is attributed to lower dispersion in this fiber.

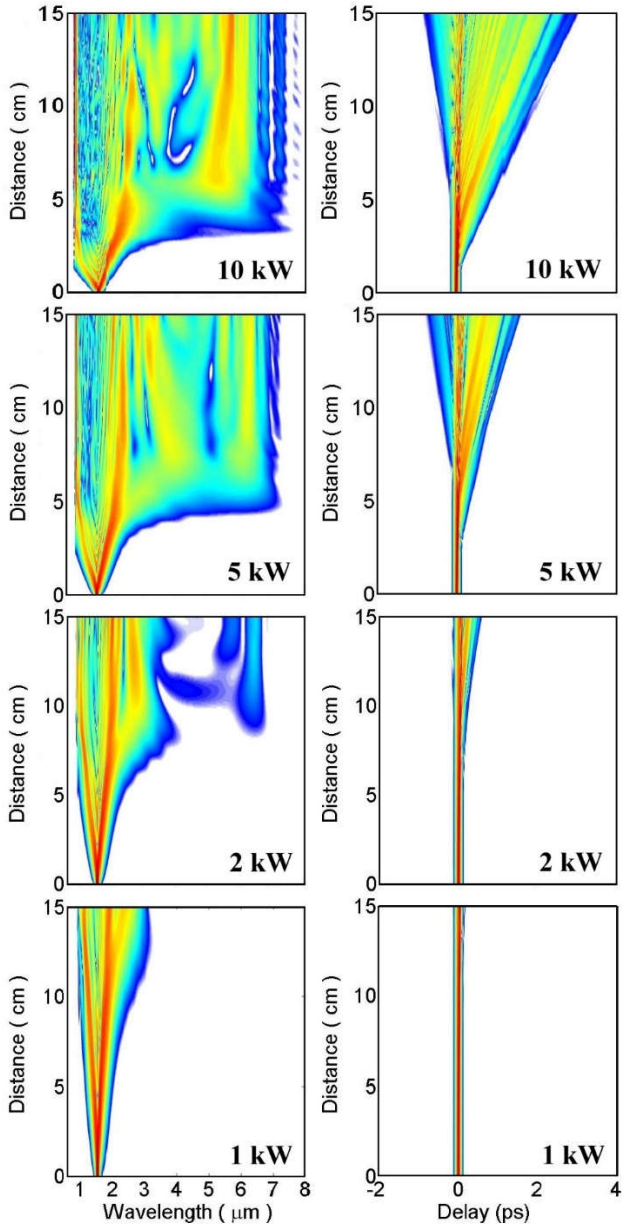

Fig. 14 Spectral and temporal evolution of supercontinuum generation by pumping 50 fs pulse in a 15 cm long PCF at different peak power.

## 5. Conclusion

We have designed a three layer highly nonlinear soft glass (SF57) photonic crystal fiber which consists of seven rings of air holes. Air hole diameter in different air hole rings are uniquely arranged to yield optimized value of nonlinearity and dispersion that is required for efficient supercontinuum generation at low pump power. With appropriate fiber parameters, the typical value of achievable nonlinearity and dispersion at the wavelength $\lambda = 1.55~\mu m$ are $\sim 1078~W^{-1}km^{-1}$ and $\sim 4~ps/nm/km$ respectively. The fiber has been employed for numerical investigation of supercontinuum generation using 50 fs pump pulses of 5 kW peak power at a wavelength 1.55 $\mu m$. Due to very small value of the dispersion coefficient, large number of higher order solitons (approximately 87 at 5 kW) generated. Self-interaction among solitons occurs due to cross-phase modulation and four-wave mixing processes. Which help to generate a long rangered-shifted DW approximately 6300 nm spanning from 900 to 7200 nm. We also investigated the compatibility of supercontinuum generation with the pulse peak power where we found that the increment in peak power increases the XPM and FWM process which disturbed the supercontinuum flatness for the continuing fiber length. To our best knowledge this is the broadest supercontinuum in lead silicate PCF employing pump pulses of 5 kW peak power.

### Acknowledgement

We thank gratefully anonymous referees for insightful comments and valuable suggestions. This work is supported by the Department of Science and Technology (DST), Government of India through the R&D grant SR/S2/LOP-17/2010.

### Reference

[1] Knight, J. C., Birks, T. A., Russell, P. S. J., & Atkin, D. M. *Optics letters*, *21*(19), 1547-1549. (1996).

[2] Khan, K. R., Bidnyk, S., & Hall, T. J. *Progress in Electromagnetics Research M*, *22*, 179-189. (2012).

[3] Sharma, M., Konar, S., & Khan, K. R. *Journal of Nanophotonics*, *9*(1), 093073-1. (2015).

[4] Bhattacharya, R., & Konar, S. *Optics & Laser Technology*, *44*(7), 2210-2216. (2012).

[5] Sharma, M., Borogohain, N., & Konar, S. *Journal of Lightwave Technology*, *31*(21), 3339-3344. (2013).

[6] Sharma, M., Borgohain, N., & Konar, S. *Physics Express4*(1), 1-9. (2014).

[7] Xing-Ping, Z., Shu-Guang, L., Ying, D., Ying, H., Wen-Qi, Z., Yin-Lan, R.,& Monro, T. M. *Chinese Physics B*, *22*(1), 014215. (2013).

[8] Tiwari, M., & Janyani, V. *Lightwave Technology, Journal of*, *29*(23), 3560-3565. (2011).

[9] Miret, J. J., Silvestre, E., & Andrés, P. *Optics express*, *17*(11), 9197-9203. (2009).

[10] Leong, J. Y., Petropoulos, P., Price, J. H., Ebendorff-Heidepriem, H., Asimakis, S., Moore, R. C., & Richardson, D. J. *Lightwave Technology, Journal of*, *24*(1), 183-190. (2006).

[11] Buczynski, R., Bookey, H. T., Pysz, D., Stepien, R., Kujawa, I., McCarthy, J. E.& Taghizadeh, M. R. *Laser Physics Letters*, *7*(9), 666. (2010).

[12] Feng, X., Mairaj, A. K., Hewak, D. W., & Monro, T. M. *Journal of lightwave technology*, *23*(6), 2046. (2005).

[13] Liao, M., Gao, W., Cheng, T., Duan, Z., Xue, X., Suzuki, T., &Ohishi, Y. *Optics express*, *20*(26), B574-B580. (2012).

[14] Kudlinski, A., George, A. K., Knight, J. C., Travers, J. C., Rulkov, A. B., Popov, S. V., & Taylor, J. R. *Optics Express*, *14*(12), 5715-5722. (2006).

[15] Kudlinski, A., Pureur, V., Bouwmans, G., &Mussot, A. *Optics letters*, *33*(21), 2488-2490. (2008).

[16] Saleh, M. F., Chang, W., Travers, J. C., Russell, P. S. J., &Biancalana, F. *Physical review letters*, *109*(11), 113902. (2012).

[17] Dudley, J. M., Genty, G., & Coen, S. *Reviews of modern physics*, *78*(4), 1135. (2006).

[18] Coen, S., Chau, A. H. L., Leonhardt, R., Harvey, J. D., Knight, J. C., Wadsworth, W. J., & Russell, P. S. J. *JOSA B*, *19*(4), 753-764. (2002).

[19] Husakou, A. V., & Herrmann, J. *JOSA B*,*19*(9), 2171-2182. (2002).

[20] Wang, Y., Zhao, Y., Nelson, J. S., Chen, Z., &Windeler, R. S. *Optics Letters*, *28*(3), 182-184. (2003).

[21] Zabusky, N. J. In *Nonlinear Partial Differential Equations: A Symposium on Methods of Solution* (pp. 223-256). Academic Press New York. (1967).

[22] Roy, S., Bhadra, S. K., & Agrawal, G. P. *Optics letters*,*34*(13), 2072-2074. (2009).

[23] Liu, X., Lægsgaard, J., Møller, U., Tu, H., Boppart, S. A., &Turchinovich, D. *Optics letters*,*37*(13), 2769-2771. (2012).

[24] Tu, H., Lægsgaard, J., Zhang, R., Tong, S., Liu, Y., &Boppart, S. A.*Optics express*, *21*(20), 23188-23196. (2013).

[25] Cristiani, I., Tediosi, R., Tartara, L., &Degiorgio, V. *Optics express*, *12*(1), 124-135. (2004).

[26] Austin, D. R., de Sterke, C. M., Eggleton, B. J., & Brown, T. G. *Optics express*,*14*(25), 11997-12007. (2006).

[27] Cheng, T., Deng, D., Xue, X., Zhang, L., Suzuki, T., &Ohishi, Y. *Photonics Journal, IEEE*, *7*(1), 1-7. (2015).

[28] Lee, J. H., van Howe, J., Xu, C., Ramachandran, S., Ghalmi, S., & Yan, M. F. *Optics letters*, *32*(9), 1053-1055. (2007).

[29] Roy, S., Ghosh, D., Bhadra, S. K., Saitoh, K., &Koshiba, M. *Applied optics*, *50*(20), 3475-3481. (2011).

[30] Li, X., Chen, W., Xue, T., Bi, W., Gao, W., Hu, L., & Liao, M. *Journal of Applied Physics*, *117*(10), 103103. (2015).

[31] White, T., McPhedran, R., Botten, L., Smith, G., & De Sterke, C. M. *Optics Express*, *9*(13), 721-732. (2001).

[32] Mogilevtsev, D., Birks, T. A., & Russell, P. S. J. *Journal of lightwave technology*, *17*(11), 2078. (1999).

[33] Shi, S., Chen, C., & Prather, D. W. *JOSA A*, *21*(9), 1769-1775. (2004).

[34] Khan, K. R., & Wu, T. X. *Selected Topics in Quantum Electronics, IEEE Journal of*, *14*(3), 752-757. (2008).

[35] Khan, K., Mahmood, M., & Biswas, A. *Selected Topics in Quantum Electronics, IEEE Journal of*(2014).

[36] Park, K. N., & Lee, K. S. *Optics letters*, *30*(9), 958-960. (2005).

[37] Xu, F., Zhang, Y., Hong, W., Wu, K., & Cui, T. J. *Theory and Techniques, IEEE Transactions on*, *51*(11), 2221-2227. (2003).

[38] Hagness, S. C., &Taflove, A. *Norwood, MA: Artech House*. (2000).

[39] BandSOLVE, RSoft Design Group Inc. USA, (2003).

[40] Gallagher, D., & Design, P. Photonic CAD matures. *IEEE LEOS NewsLetter*. (2008).

[41] Dudley, J. M., & Taylor, J. R. (Eds.). *Supercontinuum generation in optical fibers*. Cambridge University Press. (2010).

[42] Agrawal, G. P. *Nonlinear fiber optics*. Academic press. (2007).

[43] Genty, G., Lehtonen, M., &Ludvigsen, H. *Optics express*, *12*(19), 4614-4624. (2004).

[44] Skryabin, D. V., &Yulin, A. V. *Physical Review E*, *72*(1), 016619. (2005).

[45] Efimov, A., Yulin, A. V., Skryabin, D. V., Knight, J. C., Joly, N., Omenetto, F. G., & Russell, P. *Physical review letters*, *95*(21), 213902. (2005).

[46] Peleg, A., Chertkov, M., &Gabitov, I. *JOSA B*,21(1), 18-23. (2004).

[47] Monro, T. M., Kiang, K. M., Lee, J. H., Frampton, K., Yusoff, Z., Moore, R., & Richardson, D. J. High nonlinearity extruded single-mode holey optical fibers. In *Optical Fiber Communication Conference* (p. FA1). Optical Society of America. (2002, March).

[48] Tittel, F. K., Richter, D., & Fried, A. In *Solid-State Mid-Infrared Laser Sources* (pp. 458-529). Springer Berlin Heidelberg. (2003).

[49] Humbert, G., Wadsworth, W., Leon-Saval, S., Knight, J., Birks, T., St J Russell, P., &Stifter, D. *Optics express*, *14*(4), 1596-1603, (2006).

[50] Jones, D. J., Diddams, S. A., Ranka, J. K., Stentz, A., Windeler, R. S., Hall, J. L., &Cundiff, S. T., *Science*, *288*(5466), 635-639. (2000).

[51] Mürtz, M., &Hering, P., Online monitoring of exhaled breath using mid-infrared laser spectroscopy. *In Mid-Infrared Coherent Sources and Applications*Springer Netherlands. (pp. 535-555). (2008).

[52] Gong, T. R., Yan, F. P., Wang, L., Liu, P., Li, Y. F., & Jian, S. S. *Optoelectronics Letters*, *4*(2), 110-113, (2008).